\documentclass{tran-l}

\usepackage{amsmath,amsthm,amssymb,xcolor,verbatim,graphicx,url}

\newtheorem{theorem}{Theorem}[section]
\newtheorem{lemma}[theorem]{Lemma}
\newtheorem{claim}[theorem]{Claim}

\newtheorem{corollary}[theorem]{Corollary}

\newtheorem{concl}[theorem]{Conclusion}

\newtheorem{conv}[theorem]{Convention}
\newtheorem*{claim*}{Claim}

\newtheorem{fact}[theorem]{Fact}
\newtheorem{expl}[theorem]{Example}
\newcommand{\br}{\vspace{2mm}}

\newtheorem{rmk}[theorem]{Remark}
\newtheorem{definition}[theorem]{Definition}

\newtheorem{disc}[theorem]{Discussion}

\newcommand{\dom}{\operatorname{dom}}

\newcommand{\mcp}{\mathcal{P}}

\newcommand{\trv}{\mathbf{t}}
\newcommand{\mch}{\mathcal{H}}

\newcommand{\vp}{\varphi}

\newcommand{\tlfeq}{\trianglelefteq}
\newcommand{\ma}{\mathcal{A}}
\newcommand{\mct}{\mathcal{T}}
\newcommand{\rstr}{\upharpoonright}

\newcommand{\bip}{\operatorname{Bip}}

\newcommand{\Ex}{\mathop \mathbb{E}}

\newcommand{\eps}{\epsilon}

\newcommand{\X}{\mathcal{X}}

\newcommand{\A}{\mathcal{A}}
\renewcommand{\H}{\mathcal{H}}

\newcommand{\ldim}{\mathsf{Ldim}}

\newcommand{\D}{\mathcal{D}}
\newcommand{\T}{\mathcal{T}}
\newcommand{\mcm}{\mathcal{M}}

\numberwithin{equation}{section}

\begin{document}


\title{Model theory and agnostic online learning via excellent sets}


\author{M. Malliaris}
\address{Department of Mathematics, University of Chicago} 
\email{mem@math.uchicago.edu}
\thanks{MM's research partially supported by NSF CAREER 1553653 and NSF-BSF 2051825.}

\author{S. Moran}
\address{Departments of Mathematics, Computer Science, and Data and Decision Sciences, Technion.}
\email{smoran@technion.ac.il}
\thanks{SM is Robert J.\ Shillman Fellow and support by ISF grant 1225/20, by BSF grant 2018385, by an Azrieli Faculty Fellowship, by Israel PBC-VATAT, by the Technion Center for Machine Learning and Intelligent Systems (MLIS), and by the the European Union (ERC, GENERALIZATION, 101039692). Views and opinions expressed are however those of the author(s) only and do not necessarily reflect those of the European Union or the European Research Council Executive Agency. Neither the European Union nor the granting authority can be held responsible for them.
}


\date{}








\begin{abstract} 
We {use algorithmic methods from online learning to} explore some important objects at the intersection of model theory and combinatorics, 
and find natural ways that algorithmic methods can detect and explain (and improve our understanding of) 
stable structure in the sense of model theory.  
The main theorem deals with existence of $\epsilon$-excellent sets (which are key to the Stable Regularity Lemma, a theorem characterizing the appearance of irregular pairs in Szemer\'edi's celebrated Regularity Lemma). 
We prove that $\epsilon$-excellent sets exist for any $\epsilon < \frac{1}{2}$ in 
$k$-edge stable graphs in the sense of model theory (equivalently, Littlestone classes); earlier proofs had given this only for $\epsilon < 1/{2^{2^k}}$ or so.  
 We give two proofs: the first uses regret bounds from online learning, the second uses Boolean closure 
properties of Littlestone classes and sampling. We also give a version of the dynamic Sauer-Shelah-Perles lemma appropriate 
to this setting, related to definability of types. We conclude by characterizing stable/Littlestone classes as those supporting 
a certain abstract notion of majority: the proof shows that the two distinct, natural notions of majority, arising from measure and from dimension, densely often coincide.  
\end{abstract}

\maketitle


\vspace{3mm}

In the recent papers \cite{almm}, \cite{blm},  
 ideas from model theory played a role in the conjecture, and then the proof, that Littlestone classes (which model theorists would call stable) 
 are precisely those which can be 
PAC learned in a differentially private way. We direct the reader to those papers for precise statements and further literature review.   
The present work may be seen as complementary to that work in that it shows, perhaps even more surprisingly, that ideas and techniques 
can travel profitably in the other direction.  

In the introduction below, we briefly present three points of view (combinatorics, online learning, model theory) which inform this work.  
The aim is to allow the paper to be readable by people in all three communities. 
Before this, we explain the results, deferring some definitions to the introduction below. 

The technical contributions of the paper are as follows: 

\begin{enumerate}
\item   The main theorem: Theorem \ref{theorem-1}, page \pageref{theorem-1} below. 
Informally, stable classes $\equiv$ Littlestone classes have large 
$\epsilon$-excellent sets for \emph{any} $\epsilon < \frac{1}{2}$ (and this is a characterization).  

The technical significance of this result is that it extends a useful proof from \cite{MiSh:978} which required $\epsilon < \frac{1}{2^d}$ to 
essentially any $\epsilon$ for which ``$\epsilon$-excellent'' is well defined.
It is nice to have this answer, but the longer-term effect of this result, in the opinion of the authors, 
may come instead from the methods of proofs. 
In fact, we give two distinct proofs.  
The first proof, in \S \ref{s:e-r}, uses \emph{no-regret} and \emph{multiplicative-weights algorithms} 
from machine learning. These methods have a long history in computer science. 
We found it surprising that these applied rather naturally to extract model theoretic information and we wonder what else may be done with this. The second proof, in \S \ref{s:closure}, 
uses the VC theorem and closure properties of Littlestone (stable) classes under Boolean operations. There are 
some interesting quantitative questions here.

\br
\item  Theorem \ref{theorem-3}, page \pageref{theorem-3} below.   
Informally, this gives a new characterization of stable/Littlestone classes as those admitting a certain axiomatic notion of majority. 

This theorem (or perhaps its definition) resolves an apparent discrepancy between two notions of majority arising {in} this world: the 
dimensional majority used in 
 rank or Littlestone dimension, and the counting majority used in goodness and excellence.  It shows 
that the two can always be made to coincide in a precise way.  Earlier analogues of this discrepancy could be seen, for instance, in 
proofs of stable regularity which tended to 
choose either one or the other (cf.\ \cite{MiSh:978}, \cite{MiSh:E98} on the one hand and \cite{malliaris-pillay} on the other) according to 
whether the focus was on bounds or definability.   

\br
\item  Theorem \ref{dssp}, page \pageref{dssp} below.  
Informally, we can detect whether or not a class is stable (Littlestone)
by counting the number of \emph{algorithms} needed to simulate it on an unknown tree of fixed height. 

The statement is a mild extension of a known result in machine learning for Littlestone classes 
(removing the assumption of ``oblivious,'' i.e. replacing sequences with trees).  
The known result is what explains the contribution of finite Littlestone dimension in \S \ref{s:e-r}, 
so some exposition of this was necessary in order to make that proof clear to the model-theoretic reader; we took the occasion 
to prove the extension and to make clear the connection with definability of types. 

Although it is technically the simplest contribution, we feel that the discovery of a context in which 
the correct analogue of \emph{types} is \emph{algorithms} is 
significant.
\end{enumerate}

\br

\noindent {We thank the anonymous referee for many thoughtful and helpful suggestions.}

\section{Introduction}

{In this section we give a high-level overview from three different, though interconnected, points of view, and informally describe the work. 
We defer some formal definitions to later sections.}

\subsection{{Context from combinatorics}}
{Szemer\'edi's celebrated Regularity Lemma for finite graphs says essentially that any huge finite graph $G$ can be well approximated by 
a much smaller random graph.   
The lemma gives a partition of any such 
$G$ into pieces of essentially equal size 
so that edges are distributed uniformly between most pairs of pieces ({that is, most pairs are $\epsilon$-regular for some $\epsilon>0$ 
given in advance\footnote{{When $A,B$ are finite sets of vertices, 
let $e(A,B)$ denote the number of edges between $A$ and $B$, and let $d(A,B) = e(A,B)/|A||B|$ denote the density. Recall that $(A,B)$ is called $\epsilon$-regular if for all 
$A^\prime \subseteq A$ with $|A^\prime| \geq \epsilon |A|$, and all $B^\prime \subseteq B$ with $|B^\prime| \geq \epsilon |B|$, 
we have $|d(A,B) - d(A^\prime, B^\prime)| < \epsilon$.}}}).  
Szemer\'edi's original proof allowed for some pairs to be irregular, and he asked if this was necessary \cite{sz1}.  
As described in \cite[\S 1.8]{ks}, 
it was observed by several researchers including 
Alon, Duke, Leffman, R\"odl and Yuster \cite{alon} and 
Lov\'asz, Seymour, Trotter that irregular pairs are unavoidable due to the counterexample of half-graphs. 
A $k$-half graph has distinct vertices 
$a_1, \dots, a_k$, $b_1, \dots, b_k$ such that there is an edge between $(a_i, b_j)$ if and only if $i<j$.}

Malliaris and Shelah showed that half-graphs characterize the existence of irregular pairs in Szemer\'edi's lemma, by proving a  
stronger regularity lemma for $k$-edge stable graphs called the Stable Regularity Lemma \cite{MiSh:978}.  
(A graph is called $k$-edge stable if it contains no $k$-half graph. This should remind a model theoretic reader of 
the negation of the order property. 
The Stable Regularity Lemma 
says that a finite $k$-edge stable graph can be equipartitioned into $\leq m$ pieces  
(where $m$ is polynomial in $\frac{1}{\epsilon}$) such that \emph{all} pairs of pieces are regular, with densities close to $0$ or $1$. 
Two of these conditions, the improved size of the partition and the densities of regular pairs being near $0$ or $1$,  
are already expected from finite VC dimension, see \cite{alon}, \cite{lovasz-szegedy}, though here by a different proof. 
({See also section 1.3 for more on 
VC dimension, which has also a venerable history in model theory. After stable regularity, there began an active line of work looking via model theory at these questions; the introduction to \cite{tw} surveys this literature}.) 
For an exposition of stable regularity, and the model theoretic ideas behind it, see \cite{MiSh:E98}. 

A central idea in the stable regularity lemma was 
that $k$-edge stability for small $k$ means it is possible to find large ``indivisible'' sets,  
so-called \emph{$\epsilon$-excellent} sets. 

{To informally recall the definition (formal definitions will be given in \ref{d:excellence} below)}, 
let $0 < \epsilon < \frac{1}{2}$. Let $G = (V,E)$ be a finite graph. Following \cite{MiSh:978}, 
say $B \subseteq V$ is \emph{$\epsilon$-good} if for any $a \in V$, one of $\{ b \in B : (a,b) \in E \}$, 
$\{ b \in B : (a,b) \notin E \}$ has size $<\epsilon|B|$.  If the first [most $b \in B$ connect to $a$], write $\trv(a,B) = 1$, and 
if the second [most $b \in B$ do not connect to $a$] write $\trv(a,B) = 0$. 
Say that $A \subseteq V$ is \emph{$\epsilon$-excellent} if for any $B \subseteq V$ which is 
$\epsilon$-good, one of $\{ a \in A : \trv(a,B) = 1 \}$, $\{ a \in A : \trv(a,B) = 0 \}$ has size $<\epsilon|A|$. 
Informally, any $a \in A$ has a majority opinion about any $\epsilon$-good $B$ by definition of good, and 
excellence says that additionally, a majority of elements of $A$ have the same majority opinion. 
Observe that if $A$ is $\epsilon$-excellent it is $\epsilon$-good, because any set of size one is $\epsilon$-good. 

{A partition into excellent sets is quickly seen to have no irregular pairs (for a related $\epsilon$).}

{Notice that while, e.g., $\frac{1}{4}$-good implies $\frac{1}{3}$-good, the same is 
a priori not true for $\epsilon$-excellent, because the definition of $\epsilon$-excellence quantifies over $\epsilon$-good sets.  
{See Example \ref{e:ge} below.} 
For the stable regularity lemma, it was sufficient to show that large $\epsilon$-excellent sets exist in 
$k$-edge stable graphs for $\epsilon < \frac{1}{2^{2^k}}$ or so.  In this language, one contribution of the present paper is 
a new proof for existence of $\epsilon$-excellent sets in $k$-edge stable graphs, which works for any $\epsilon < \frac{1}{2}$, i.e. any $\epsilon$ for which excellence is well defined.}

\br
\subsection{{Context from online learning}} \label{c:online-learning}
{Online learning is a well-studied model for algorithms making real-time predictions on sequentially arriving data.
In the basic version of online learning, a game between a learner and an adversary is played sequentially for $T$ rounds as follows. At each stage $t=1,\ldots, T$ the adversary presents the learner with a point $x_t\in X$ of the adversary's choosing and asks the learner to predict the label $y_t\in\{0,1\}$. Then, the learner makes a prediction $\hat y_t$ after which the correct label $y_t$ is revealed to the learner.\footnote{Note that $\hat y_t=\hat y_t(x_1,\ldots, x_t)$ may depend on everything that happened up to round $t$. {When the learner uses randomness to make predictions, the adversary isn't able to see the model's internal randomness. Thus, the learner shares only the probability \(p_t\) that the predicted outcome \(\hat{y}_t\) will be 1. The actual prediction \(\hat{y}_t\) is made randomly, based on this probability, and only after the adversary has chosen \(y_t\).}} The learner receives a penalty of $1$ for a mistake.
The adversary's goal is to play so as to maximize the number of mistakes, and the learner's goal is to minimize them. The performance of the learner is measured in terms of its \emph{regret}:
\[\sum_{t=1}^T 1[\hat y_t\neq y_t] - \min_{h\in \mch}\sum_{t=1}^T 1[ h(x_t)\neq y_t]\]
that is, the excess number of mistakes the learner makes compared to the best function $h\in\mch$. Say $\mch$ is (agnostic) online learnable if the learner has an algorithm whose regret against any adversary is sublinear in $T$. An important special case is online learning in the realizable case, in which the adversary is restricted to produce sequences $\{(x_t,y_t)\}_{t=1}^T$ that are realizable by (or consistent with) the class $\mch$: that is, for which there exists $h\in \mch$ such that $h(x_t)=y_t$ for all $t=1,\ldots, T$. Notice that in the realizable case, the regret of the learner specializes to the absolute number of mistakes the learner made on the input sequence.}

The online learning setting shifts the basic context from graphs to \emph{hypothesis classes}, i.e. pairs $(X, \mch)$ where 
$X$ is a finite or infinite set and $\mch \subseteq \mcp(X)$ is a set of subsets of $X$, called {hypotheses} or predictors.   We will identify
elements $h \in \mch$ with their characteristic functions, and write ``$h(x) = 1$'' for ```$x \in h$'' and $h(x) =0$ otherwise. 
Any such hypothesis class can be naturally viewed as a bipartite graph on the disjoint sets of vertices $X$ and $\mch$ 
with an edge between $x \in X$ and $h \in \mch$ if and only if $h(x) = 1$.   
However, something which {may be} 
 lost in this translation is a powerful understanding in the computer science community of the role of  
dynamic/adaptive/predictive arguments.   This perspective is an important contribution to the 
proofs below, and seems to highlight some understanding currently missing in the other contexts.

{Consider the idea of a mistake tree:
we have a full binary tree whose internal nodes are labeled by elements of~$X$, and which is realized by $\mch$ in the following sense. 
We can think of the process of 
traversing a root-to-leaf path, that is, a branch, in a tree of height $d$ as being described by a sequence of pairs $(x_i, y_i) \in X \times \{ 0, 1 \}$, 
for $1 \leq i \leq d$, recording that at 
step $i$ the node we are at is labeled by $x_i$ and we then travel right (if $y_i = 1$) or left (if $y_i = 0$) to a node labeled by $x_{i+1}$, 
and so on.  Say that $h \in \mch$ realizes a given branch $(x_1, y_1), \dots (x_d, y_d)$ if $h(x_i) = y_i$ for all 
$1 \leq i \leq d$, and say that a given tree is shattered by $\mch$ if each branch is realized by some $h \in \mch$. Call such a tree 
a \emph{mistake tree of height $d$} for $\mch$. 
(For a more extensive discussion see \cite{ML-book} \S 18.1-18.2.)}

The Littlestone dimension $d$ of~$\mch$, denoted~$\ldim(\mch)$, is the depth of the largest mistake tree, i.e., the largest complete [binary] tree that is shattered by $\mch$, or $\infty$ if no such bound exists. 
Notice that existence of such a tree helps the adversary force at least $d$ mistakes in the realizable case.  
$\mch$ is called a Littlestone class if it has finite Littlestone dimension; for reasons explained in the next subsection, we may prefer to say that $(X, \mch)$ is a Littlestone pair.   Littlestone~\cite{littlestone88} and Ben-David, P{\'{a}}l, and Shalev-Shwartz~\cite{ben-david09agnostic} proved that $\ldim$ characterizes online learnability of the class. Quantitatively, the optimal regret is equal to $\Theta(\sqrt{\ldim\cdot T})$ (\cite{ben-david09agnostic,alon21}) and in the realizable case the optimal mistake bound is $\ldim$~\cite{littlestone88}.\footnote{{More precisely, $\ldim$ is the optimal bound achievable in the realizable setting by \emph{deterministic} learners. The optimal bound achievable by randomized learner is characterized by the randomized Littlestone dimension~\cite{Mehalel23}.}}

\subsection{Context from model theory}

Consider again the case of a finite bipartite graph $G$ with vertex set~$X \cup Y$ and edge relation $R$ (abstracting the study of 
a formula $\vp(\bar{x}, \bar{y})$). Following logical notation we write 
$R(a,b)$ or $\neg R(a,b)$ to denote an edge or a non-edge. Define a [full] \emph{special tree} \label{tree-page} of 
height $n$ to have    
 internal nodes $\{ a_\eta : \eta \in {^{n>}2} \}$ from $X$ and indexed by binary sequences of length $<n$ and leaves 
$\{ b_\rho : \rho \in {^n 2} \}$ from~$Y$ and indexed by binary sequences of length exactly $n$, which satisfy the following.\footnote{On this notation, see Convention \ref{conv:notation}.}  
For any $a_\eta$ and $b_\rho$, if $\eta$ is an initial segment of $\rho$ (notation: $\eta \tlfeq \rho$), then $R(a_\eta, b_\rho)$ if 
${\eta^\smallfrown \langle 1 \rangle} \tlfeq \rho$ and $\neg R(a_\eta, b_\rho)$ if ${\eta^\smallfrown \langle 0 \rangle} \tlfeq \rho$. 
A key ingredient in the proof of the Stable Regularity Lemma was the following special case of Shelah's Unstable Formula Theorem 
\cite[II.2.2]{shelah}.  For a bipartite graph $G$ as above, 
if $G$ has a full special tree of height $n$, then it has a half-graph of size about $\log n$, with the $a$'s chosen from $X$ and the 
$b$'s chosen from $Y$ $($i.e., it is not $k$-edge stable in the sense above$)$. Moreover, 
if $G$ has a half-graph of size $k$, it has a full special tree of height about $\log k$. (These bounds are due to Hodges, {see \cite{hodges1} or \cite{hodges} Lemma 6.7.9, and see \S \ref{s:open-problems} \# 2 below.})   

It was noticed by Chase and Freitag \cite{cf} that the condition of model-theoretic stability (Shelah's 2-rank; 
e.g., in this language, no full special tree of height $n$ for some finite $n$) 
corresponds to finite Littlestone dimension, and 
they used this to give natural examples of Littlestone classes using stable theories.  
An analogous connection 
between so-called \emph{NIP theories} and VC dimension had 
also been previously observed by Laskowski \cite{laskowski} and led to results in learning theory, particularly in the context of compression schemes, see for instance Livni and Simon \cite{livni-simon}, and on related definability questions, see for instance 
Eshel and Kaplan \cite{eshel-kaplan}, but also some of the first polynomial bounds for VC dimension for sigmoidal neural networks, see Karpinski and Macintyre \cite{karpinski-macintyre}. 

The following discussion reflects an understanding developed in the online learning papers Alon-Livni-Malliaris-Moran \cite{almm}, Bun-Livni-Moran \cite{blm}, where 
model theoretic ideas had played a role in the proof that the Littlestone classes are precisely those which can be PAC-learned in a 
differentially private way. One contribution of the model theoretic point of view for online learning in general, and for our present argument in particular, is that a condition in online learning which appears inherently asymmetric, 
namely the Littlestone dimension (it treats elements and hypotheses as different kinds of objects; they play fairly different roles in the 
partitioning) is equivalent to a condition which is extremely symmetric, namely existence of half-graphs (when switching the roles of 
$X$ and $Y$ in a half-graph, it suffices to rotate the picture). Thus if $\mch$ is a Littlestone class, 
the ``dual'' class obtained by setting $X^\prime = \mch$ and $\mch^\prime = \{ \{  h \in \mch : h(x) = 1  \}  : x \in X \}$ is also. 
In online learning, $k$-edge stability also has a natural meaning: Threshold dimension $k$, that is, there do not exist elements 
$a_1, \dots, a_k$ from $X$ and hypotheses $h_1, \dots h_k$ from $\mch$ such that $h_j(a_i) = 1$ if and only if $i<j$. 
In what follows, we sometimes refer to $(X, \mch)$ as a \emph{Littlestone pair}, rather than simply saying that $\mch$ is a 
Littlestone class, to emphasize this line of thought.

\br
\section{Prior results and a characterization} \label{s:prior}

\begin{conv}
Following convention, we say ``$(X, \mch)$ is a hypothesis class''  to mean that $X$ is a set and $\mch$ is a set of subsets of $X$ 
$($sometimes identified with their characteristic functions$)$.   
\end{conv}

In the language of online learning, $(X, \mch)$ is a Littlestone class if it has finite Littlestone dimension $(\operatorname{Ldim})$ 
in the sense of \cite{ML-book} Chapter 21. A combinatorial or model theoretic reader may take the Littlestone dimension simply 
to be the maximal height of a special tree (\S 1.3 above), an instance of Shelah's 2-rank, see \cite{cf} or \cite{onshuus}.  

The original facts about good and excellent sets in \cite{MiSh:978} were proved for $k$-edge stable graphs (\S 1.1 above); 
the translation to Littlestone classes is immediate, but we record this here for completeness.

\begin{definition} \label{d:excellence}
Let $0 < \epsilon < \frac{1}{2}$ and let $(X, \mch)$ be a hypothesis class. 

\begin{enumerate}
\item[(1)] Say 
$B \subseteq X$ is \emph{$\epsilon$-good} if for any $h \in \mch$, one of $\{ b \in B : h(b) = 1 \}$, 
$\{ b \in B : h(x) = 0 \}$ has size $<\epsilon|B|$.  
Write $\trv(h,B) = 1$ in the first case, $\trv(h,B) = 0$ in the second.

\item[(2)]
Say that $H \subseteq \mch$ is \emph{$\epsilon$-excellent} if for any $B \subseteq X$ which is 
$\epsilon$-good, one of  $\{ h \in H : \trv(h,B) = 1 \}$, $\{ h \in H : \trv(h,B) = 0 \}$ has size $<\epsilon|H|$. 
Write $\trv(H,B) = 1$ or $\trv(H,B) = 0$ to record this. 

\item[(3)] Define ``$H$ is an $\epsilon$-good subset of $\mch$'' and ``$A$ is an $\epsilon$-excellent subset of $X$'' in the parallel way 
switching the roles of $X$ and $\mch$. 
\end{enumerate}
\end{definition}

\begin{rmk} \label{d:monot}
\emph{The definition of $\epsilon$-good is monotonic in $\epsilon$: it becomes weaker as $\epsilon$ increases $($below $\frac{1}{2}$$)$.  
This is a priori not the case for excellence: as $\epsilon$ increases,  
the $\epsilon$-good sets $B$ quantified over may increase, {as the following illustrates.} }
\end{rmk}

{
\begin{expl} 
We first give an example of a set which is $\epsilon$-good but not $\epsilon$-excellent. 
Let $A = \{ a_1, a_2, a_3, a_4, a_5 \}$ and $B = \{ b_1, b_2, b_3, b_4, b_5 \}$ and consider the bipartite graph with vertex set 
$A, B$. Let $\epsilon$ be slightly larger than $1/5$, say, $\epsilon = 13/60$.  
Suppose 
the edges are given as follows:
\begin{itemize}
\item $a_1 \sim b_1$, and $a_1 \not\sim b_2, b_3, b_4, b_5$, so $\trv(a_1,B) = 0$.
\item $a_2 \sim b_2$, and $a_2 \not\sim b_1, b_3, b_4, b_5$, so $\trv(a_2,B) = 0$. 
\item $a_3 \not\sim b_3$, and $a_3 \sim b_1, b_2, b_4, b_5$, so $\trv(a_3,B) = 1$. 
\item $a_4 \not\sim b_4$, and $a_4 \sim b_1, b_2, b_3, b_5$, so $\trv(a_4,B) = 1$.
\item $a_5 \not\sim b_5$, and $a_5 \sim b_1, b_2, b_3, b_4$, so $\trv(a_5,B) = 1$. 
\end{itemize}
Then $A$ and $B$ are both $\epsilon$-good but since 
$(1-\epsilon)|A| > |\{ a \in A : \trv(a,B) = 1 \}| > \epsilon|A|$, i.e. $1-\epsilon > 2/5 > \epsilon$, 
$A$ is not $\epsilon$-excellent $($nor is $B$ for the parallel reason$)$.  
\end{expl}

{
\begin{expl}
Let $\epsilon_1$ be slightly above $1/5$, say 
{$\epsilon_1 = 13/60$}, and let $\epsilon_2$ be slightly above $1/3$, say $21/60$. 
Next we give an example of a set $B$ which 
$(\star)$ is 
$\epsilon_2$-good, but no $B^\prime \subseteq B$ with $|B^\prime| > 1$ is $\epsilon_1$-good.

Let $B$ have vertices $b_0, b_1, b_2, b_3, b_4, b_5$. Suppose that: 
\begin{itemize}
\item every vertex in the graph either connects to at most two vertices of $B$, or 
connects to all but at most two vertices of $B$.

\item for every two distinct vertices of $B$ 
there is an element of the graph which 
connects only to them. 
\end{itemize}
To verify $(\star)$ note that if $B^\prime \subseteq B$ has size 2 or 4 it can be split in half, if it has size 3 or 6 it can be split 1/3, 2/3, and if it has size 5 it can be split 2/5, 3/5. So $B^\prime$ is not $\epsilon_1$-good unless it is a singleton.  
On the other hand, $B$ is clearly $\epsilon_2$-good. 
\end{expl}

\begin{expl} \label{e:ge}
To find an example where excellence changes with $\epsilon$, we combine the two previous examples.

Let $A = \{ a_1, a_2, a_3, a_4, a_5 \}$ 
and let $B = \{ b_0, b_1, b_2, b_3, b_4, b_5 \}$. Consider the bipartite graph with vertex set $A \cup C$ on one side and $B$ on the other, with edges given according to the following pattern:

\begin{itemize}
\item $a_1 \sim b_0, b_1$, and $a_1 \not\sim b_2, b_3, b_4, b_5$.
\item $a_2 \sim b_2$, and $a_2 \not\sim b_0, b_1, b_3, b_4, b_5$.  
\item $a_3 \not\sim b_3$, and $a_3 \sim b_0, b_1, b_2, b_4, b_5$.  
\item $a_4 \not\sim b_4$, and $a_4 \sim b_0, b_1, b_2, b_3, b_5$. 
\item $a_5 \not\sim b_5$, and $a_5 \sim b_0, b_1, b_2, b_3, b_4$.  
\item For every pair of distinct elements of $B$, there is an element of $C$ which connects to them and to no other elements of $B$.
\item There are no other edges.  
\end{itemize}
Let $\epsilon_1 = 13/60$ and 
$\epsilon_2 = 21/60$. $A$ is $\epsilon_1$-good and thus $\epsilon_2$-good. Also, $A$ is $\epsilon_1$-excellent because the 
only $\epsilon_1$-good subsets of $B$ are the singletons. However $B$ is $\epsilon_2$-good, 
and so witnesses that $A$ is not $\epsilon_2$-excellent. 
\end{expl}
}

Below we use the possible ``gaps'' in excellence arising from nonmonotonicity as a 
motivation to give new proofs and new theorems.\footnote{{It also may be interesting to investigate the 
limits of these gaps. Added in revision: note that in the course of significant recent work on hypergraphs, 
\cite[Corollary 5.10]{tw} Terry and Wolf consider the relation of goodness and \emph{regularity}, 
showing that a pair of $\epsilon$-good sets is essentially $\sqrt{\epsilon}$-regular, which 
Malliaris and Shelah \cite[Claim 5.17]{MiSh:978} show for $\epsilon$-excellent sets.}}   

\br
The key point about excellent sets that they exist characteristically in stable (Littlestone) classes:  
}

\begin{fact}[\cite{MiSh:978} Claim 5.4 or \cite{MiSh:E98} Claim 1.8, in our language] \label{fact:978}
Let $(X, \mch)$ be a Littlestone class, $\ldim(\mch) = d$ and $0 < \epsilon < \frac{1}{2^d}$. 
For any finite $H \subseteq \mch$ there is $A \subseteq H$, $|A| \geq \epsilon^d |H|$ such that 
$A$ is $\epsilon$-excellent. 
\end{fact}

The proof of Fact \ref{fact:978} proceeds by noting that if 
$H = H_\emptyset$ is not $\epsilon$-excellent, then there is some $\epsilon$-good $A = A_\emptyset$ which witnesses this failure, 
splitting $H_\emptyset$ naturally into $H_{\langle 0 \rangle}$ and $H_{\langle 1 \rangle}$ according to $\trv$.  
If either of these is excellent, we stop; if not, continue inductively  
to label the internal nodes and leaves of a full binary tree with $A$'s and $H$'s respectively. Suppose we arrive to height $d$. 
To extract a Littlestone tree, or equivalently a full special tree 
(p. \pageref{tree-page} above), choose a $h_\rho$ from each $H_\rho$
and then show it is possible to choose a suitable  
$a_\eta$ from each $A_\eta$ by using $\epsilon < 2^{-d}$, the definition of good (really, of $\trv$) 
and the union bound: {that is, for each given $\eta$, each $h_\rho$ with $\rho$ extending $\eta$ rules out at most an $\epsilon$-fraction of elements of $A_\eta$.
So  by the choice of $\epsilon$, there is a remaining element in $A_\eta$ acceptable to all such $h_\rho$.}\footnote{{Observe that 
this appeal to the union bound won't work for Example \ref{e:ge} above.}} 

{Since finding such a Littlestone tree contradicts $\ldim(\mch)=d$, it must be that for some $\rho$ of length $<d$, $H_\rho$ is excellent. }  

Note moreover that the same proof works to show existence of $\epsilon$-good sets, simply by taking the sets $A$ to be singletons 
(note that any singleton is trivially $\epsilon$-good). In this case, the union bound is not needed and the proof works for any $\epsilon < \frac{1}{2}$.  

\begin{fact}[\cite{MiSh:978}, see above] \label{c:good}  
Let $(X, \mch)$ be a Littlestone class, $\ldim(\mch) = d$, $0 < \epsilon < \frac{1}{2}$. 
For every finite $A \subseteq \mch$ there is $B \subseteq A$, 
$|B| \geq \epsilon^{d}|A|$ such that $B$ is $\epsilon$-good. 
\end{fact}

{
\begin{definition} \label{d:dual}
Let $(X, \mch)$ be a hypothesis class.
Define the \emph{dual} hypothesis class to be $(X^\prime, \mch^\prime)$ where 
$X^\prime = \mch$ and $\mch^\prime = \{ \{  h \in \mch : h(x) = 1  \}  : x \in X \}$. 
\end{definition}

Definition \ref{d:dual} is perhaps most natural in the language of graphs:

\begin{definition} \label{d:bip} A hypothesis class $(X, \mch)$ gives rise to a bipartite graph 
$\bip(X, \mch)$ 
with an edge between $x$ and $h$ if and only if $h(x) = 1$.
\end{definition} 

Likewise any bipartite graph gives rise to a hypothesis class after we 
specify which side is the domain and which side is the hypotheses. The dual class $(X^\prime, \mch^\prime)$ arises from the same bipartite graph 
$\bip(X, \mch)$ but with the opposite specification. 
}

We conclude this section by observing in Claim \ref{c:equiv} that 
existence of large good sets (both in $X$ and in $\mch$) is characteristic of Littlestone classes.   
At this point, a similar result 
for excellence could also be stated, for all sufficiently small $\epsilon$.

\begin{claim} \label{c:equiv} The following are equivalent for any hypothesis class $(X, \mch)$.
\begin{enumerate}
\item For every $\epsilon < \frac{1}{2}$ there is a constant $c= c(\epsilon) > 0$ such that for every finite $A \subseteq \mch$ there exists $B \subseteq A$, 
$|B| \geq c|A|$ such that $B$ is $\epsilon$-good. 

\item For some $\epsilon < \frac{1}{2}$ and some constant $c > 0$, for every finite $A \subseteq \mch$ there exists $B \subseteq A$, 
$|B| \geq c|A|$ such that $B$ is $\epsilon$-good. 

\item $\mch$ is a Littlestone class. 

{ 
\item For some finite $k$, $\bip(X, \mch)$ does not contain any $k$-half graph as an induced subgraph.\footnote{{Such graphs are called $k$-stable graphs.}} 

\item For some finite $\ell$, $\bip(X^\prime, \mch^\prime)$ does not contain any $\ell$-half graph as an induced subgraph. 

\item The dual class $(X^\prime, \mch^\prime)$ $($in the sense of $\ref{d:dual}$$)$ is a Littlestone class. 

}
\item For every $\epsilon < \frac{1}{2}$ there is a constant $c= c(\epsilon) > 0$ such that for every finite $A \subseteq X$ there exists $B \subseteq A$, 
$|B| \geq c|A|$ such that $B$ is $\epsilon$-good. 

\item For some $\epsilon < \frac{1}{2}$ and some constant $c > 0$, for every finite $A \subseteq X$ there exists $B \subseteq A$, 
$|B| \geq c|A|$ such that $B$ is $\epsilon$-good. 
\end{enumerate}
\end{claim}

{
\begin{proof}
(1) implies (2) is immediate, and (3) implies (1) is Fact \ref{c:good}. 

To show (2) implies (3), suppose we are given $\epsilon$ and $c$ from (2).  Since any finite hypothesis class is necessarily Littlestone, we may assume 
$X$ is infinite. Choose $n$ large enough so that 
$\lfloor \epsilon \lfloor cn \rfloor \rfloor \geq 1$ and so that for any $k \geq \lceil \epsilon n \rceil$, we have 
$\min \{ \lfloor \frac{k}{2} \rfloor -1, \lfloor \frac{k-1}{2} \rfloor \} \geq \epsilon k$. 
If $\mch$ is not a Littlestone class, then we know it has infinite (i.e. not finite) Threshold dimension and so for our chosen $n$, 
there are elements $\{ x_i : i < n \}$ from $X$ and $H := \{ h_j : j < n \}$ from $\mch$ such that 
$x_i \in h_j$ if and only if $i<j$. But for any $H^\prime \subseteq H$ of size $k \geq cn$, we can pick out all the cuts of $H^\prime$
using the $x_i$'s. In particular, if $H^\prime = \{ h_{i_\ell} : \ell < k \}$, let $m = \lceil \epsilon k \rceil$ (by choice of $n$, this 
is not larger than whichever of $k/2$ or $(k-1)/2$ is an integer). 
Then $x_m$ partitions $H^\prime$ into $\{ h_{i_\ell} : 0 \leq \ell < m \}$ and $\{ h_{i_\ell} : m \leq \ell < k \}$, both of which have size 
$\geq \epsilon k$, contradicting its being $\epsilon$-good. 

The equivalence of (3) and (4) is by Shelah's unstable formula theorem, as explained in \S 1.3 above. 

The equivalence of (4) and (5) is simply because a bipartite graph $(A, B)$ contains an induced $k$-half graph if and only if 
the bipartite graph $(B, A)$ does.\footnote{{Thus to repeat the point made in \S 1.3, in connecting the property of Littlestone, 
which is not obviously self-dual, with half-graphs which are, the Unstable Formula Theorem allows us to immediately conclude 
$\mch$ is Littlestone if and only if its dual is, though of 
course with possibly different Littlestone dimensions.}} 

The remainder of the proof then goes by a parallel argument. 
\end{proof}

\begin{rmk}
Attentive readers may wonder about in the bounds in $\ref{c:equiv}$ relating mistake trees and half-graphs. 
The known finite bounds, tracing back to Hodges \cite{hodges1}, can be stated in combinatorial language as: 
from a tree of height $d$ one obtains a half-graph of size about $\log d$, and from a half-graph of length $k$ one obtains a tree of 
height about $\log k$. This is the subject of open problem $\ref{hodges-bounds}$ in $\S \ref{s:open-problems}$ below. 
\end{rmk}
}

\begin{conv} \label{conv:notation}
We clarify some notational points which hopefully will not cause confusion if explicitly pointed out.
The word ``label'' in online learning, and in this paper, usually refers to a value such as $0$ or $1$ attached to an element of $X$ 
$($say, the value of some partial characteristic function$)$. 
Nonetheless, we write ``$X$-labeled tree'' to mean a tree in which we associate to each node an element of $X$.
In set theoretic notation,  
for each integer $n$, $n = \{ 0, \dots, n-1 \}$. Also, $^x y$ denotes the set of functions from 
$x$ to $y$, as distinguished from $y^x$ which is the \emph{size} of the set of functions from $x$ to $y$.
For logicians, a tree of height $T$ has levels $0$ to $T-1$, whereas in online learning 
the same tree would have levels $1$ to $T$. 
\end{conv}

\br

\section{Existence via regret bounds} \label{s:e-r}

The aim of this section is to prove the following theorem, using regret bounds in online learning.  
{(As noted, this improves \cite{MiSh:978}, Claim 5.4 by allowing for $\epsilon < \frac{1}{2}$ rather than 
$\epsilon < \frac{1}{2^d}$, however,  when $\epsilon < \frac{1}{2^{d}}$ that proof obtains $\epsilon^d$ as the constant $c$, 
which is not obtained by the methods here.)} 

\begin{theorem} \label{theorem-1}
 \label{c:one-half} 
 For every $\epsilon < \frac{1}{2}$ and positive integer $d$ there is a constant $c = c(d, \epsilon)$ 
 such that if $\ldim(\mch) = d$ 
 and $H \subseteq \mch$ is finite, then there is 
 $A \subseteq H$, $|A| \geq \epsilon^c |H|$ such that $A$ is $\epsilon$-excellent. Moreover $c$ is upper bounded by $d_\epsilon$ 
 from $\ref{c:approxldim}$ below. 
\end{theorem}

The key ingredient in the proof is Theorem \ref{thm:regret}. 
To begin we re-present the definitions of 
good and excellent in the language of probability. 

{
\begin{disc} \label{d:motiv}
Although to our knowledge new, $\ref{d:ge}$ is a natural opening move in our context.   
For instance, it allows for an extension of the existing model theoretic definitions into the {randomized online learning} setting 
where learner and adversary are possibly playing distributions, as explained $\S \ref{c:online-learning}$ above.

In order to reason about distributions 
we need to define an appropriate probability space.
For the sake of simplicity in the present paper we focus on the case when both $\X$ and $\H$ are finite or countable. 
This allows us to use the trivial sigma algebra which consists of the entire power set, and hence avoid stating (standard) measure theoretic assumptions. 
When applying these results in the uncountable setting the 
reader is cautioned to also verify the (standard) 
measure theoretic assumptions inherited from the quoted results 
on bounds.  
\end{disc}
}

\begin{definition}[$\eps$-Good and $\eps$-Excellent Distributions] \label{d:ge} 
Let $\H\subseteq \{0,1\}^\X$. 
\begin{enumerate}
\item
We say a distribution $P$ over $\X$ is $\eps$-good w.r.t $\H$ if  
\[(\forall h\in \H): \Pr_{x\sim P}[h(x)=1]\in [0,\epsilon]\cup[1-\epsilon,1]. \]
\item Similarly, a distribution $Q$ over $\H$ is $\eps$-good if  
\[(\forall x\in \X): \Pr_{h\sim Q}[h(x)=1]\in [0,\epsilon]\cup[1-\epsilon,1].\]
\item Next, a distribution $P$ over $\X$ is $\eps$-excellent if  
\[(\forall \text{ $\eps$-good } Q ): \Pr_{h\sim Q, x\sim P}[h(x)=1]\in [0,\epsilon]\cup[1-\epsilon,1].\]
\item Finally, a distribution $Q$ over $\H$ is $\eps$-excellent if  
\[(\forall \text{ $\eps$-good } P ): \Pr_{h\sim P, x\sim Q}[h(x)=1]\in [0,\epsilon]\cup[1-\epsilon,1].\]
\end{enumerate}
\end{definition}

Distributions which trivially satisfy \ref{d:ge}(1),(2) by concentrating on a single point exist in any hypothesis class. 
In Littlestone classes, one source of nontrivial examples comes from choosing some finite 
$\epsilon$-good set $A$ and taking a distribution which assigns measure $0$ to the complement of 
$A$ and is uniform on $A$. So indeed \ref{d:ge} naturally extends the usual notion of excellent and good: 

\begin{conv}
We say that a subset of $\H$ or of $\X$ is $\eps$-good ($\eps$-excellent) if the uniform distribution
on it is $\eps$-good ($\eps$-excellent).
\end{conv}

\begin{conv}
In the process of extracting large $\eps$-excellent subsets of $\H$ 
we use full binary trees whose nodes are labelled by $\eps$-good distributions;
let us refer here to such trees as \emph{$\eps$-good trees}.
\end{conv}

\begin{definition}
{Observe that each hypothesis $h \in \mch$ naturally realizes a branch in an $\eps$-good tree $\mct$.  
An $\epsilon$-good tree $\mct$ is said to be \underline{shattered} by $\H$ if every branch  
is realized by some $h\in \H$.}
\end{definition}

We now state the key technical result of the section. 

\begin{theorem}\label{thm:regret}
Let $\H$ be a hypothesis class, let $\mathcal{T}$ be an $\eps$-good complete binary tree that is shattered by $\H$, and let $T$ denote the depth of $\T$.
Then, for every online learning algorithm~$\A$, the tree $\mathcal{T}$ witnesses
a lower bound on the regret of $\A$ {$($relative to the set of experts $\H$$)$} in the following sense.
There exist distributions $\D_1,\ldots, \D_T$ over $X\times\{0,1\}$
such that an independent sequence of random examples $(x_t,y_t)\sim \D_t, t=1,\ldots,T$ satisfies the following:
\begin{itemize}
    \item The expected number of mistakes $\A$ makes on the random sequence is at least $\frac{T}{2}$.
    \item $\exists h\in \H$ whose expected number of mistakes on the random sequence is at most $\eps\cdot T$.
\end{itemize}
Thus, the expected regret of $\A$ w.r.t $\H$ on the random sequence is at least $(\frac{1}{2}-\eps)\cdot T$.
\end{theorem}

    Before we prove this theorem, let us demonstrate how one can use
    it to bound the maximum depth of an $\eps$-good 
    tree which is shattered by a Littlestone class $\H$.    
    A central line of work in the subject has established that 
    that every class $\H$ and for every $T\in\mathbb{N}$ there exists an algorithm $\A$ whose expected\footnote{The algorithm $\A$ is randomized.} 
    regret w.r.t any sequence of examples $(x_1,y_1),\ldots, (x_T,y_T)$ is
    \begin{equation} \label{eq-ldim}
        O\bigl(\sqrt{d\cdot T}\bigr),
    \end{equation}
    where $d$ is the Littlestone dimension of $\H$, and the big oh notation conceals a fixed numerical constant.\footnote{The derivation of the (optimal) bound of $O(\sqrt{d\cdot T})$ is somewhat involved~\cite{alon21}, however a slightly weaker bound of $O(\sqrt{d\cdot T \log T})$ can be proven using elementary arguments~\cite{ben-david09agnostic}.}
    Thus, by Theorem~\ref{thm:regret}, it follows that if there exists a (complete) $\eps$-good
    tree that is shattered by $\H$ of depth $T$ then $T$ must satisfy the following inequality:
    \[\Bigl(\frac{1}{2}-\eps\Bigr)\cdot T \leq O\Bigl(\sqrt{d\cdot T}\Bigr).\]
    Indeed, the LHS in the above inequality is a lower bound on the expected regret of $\A$, where as the RHS is an upper bound on it. A simple arithmetic manipulation yields that $T=O(d/(1/2 - \eps)^2)$. 
    Thus, we get the following corollary:
    \begin{corollary}\label{c:approxldim}
    Let $\H$ be a class with Littlestone dimension $d<\infty$ and let $\eps\in \bigl[0,\frac{1}{2}\bigr]$. Denote by $d_\eps$
    the maximum possible depth of a complete $\eps$-good tree which is shattered by $\H$. (Note that $d_0=d$.)
    Then,
    \begin{equation*}
        d_\eps = O\Bigl(\frac{d}{\bigl(\frac{1}{2}-\eps\bigr)^2}\Bigr).
    \end{equation*}
    \end{corollary}
    
\begin{proof}[Proof of Theorem~\ref{thm:regret}]
Let $\T$ be a tree and $\A$ be an online algorithm as in the premise of the theorem.
    We begin with defining the distributions $\D_t$.
    We first note that the label in each distribution $\D_t$ is deterministic;
    that is, there exist a distribution $D_t$ over $X$ and a label $y_t\in\{0,1\}$
    such that a random example $(x,y)\sim \D_t$ satisfies that $y=y_t$ always (with probability $=1$) and $x_t\sim D_t$. 
    The distributions $D_i$ and labels $y_i$ correspond
    to a branch of~$\T$ as follows:
\begin{itemize}
    \item Initialize $t=1$, set the ``current'' node $v_t$ to be the root of the tree.
    \item For $t=1,\ldots, T$
    \begin{enumerate}
        \item Let $D_t$ denote the $\eps$-good distribution $D_{v_t}$ which is associated with $v_t$.
        \item Define the label $y_t$ to be $1$ if and only if 
        \[\Pr_{(x_i)_{i=1}^t \sim \prod_{i=1}^t D_i, \A}\Bigl[\A\bigl(x_t; (x_{t-1},y_{t-1}),\ldots, (x_1,y_1)\bigr)=1\Bigr]\leq 1/2,\]
        where $\A$ is the given online algorithm.
        (Note that the above probability is taken w.r.t the sampling of the $x_i$'s, as well as the randomness of $\A$ in case it is a randomized algorithm.)
        I.e.\ the adversary forces that the algorithm errs with probability at least $1/2$ on $x_t$ when given an input sequence $(x_1,y_1),\ldots, (x_{t-1},y_t), x_t$, where the $x_i$'s are sampled from the $D_i$'s.
        \item Set $v_{t+1}$ to be the root of the subtree corresponding to the label $y_t$.
    \end{enumerate}
    \item Output the sequence $(D_1,y_1),\ldots, (D_T, y_T)$.
\end{itemize}

Let $(x_t)_{t=1}^T \sim \prod_{t=1}^T D_t$ and fix $t\leq T$.
    Let $\hat y_t = \A(x_t; (x_{t-1},y_{t-1}),\ldots, (x_1,y_1)\bigr)$ be the prediction of $\A$ on $x_t$. Thus, by construction $\hat y_t\neq y_t $ with probability at least $1/2$, and therefore, by linearity of expectation:
    \[\Ex_{(x_t)_{t=1}^T \sim \prod_{t=1}^T D_t, \A}\Bigl[\sum_{t=1}^T 1[y_t\neq \hat y_t]\Bigr]\geq \frac{T}{2}.\]

It remains to show that there exists $h\in \H$ whose expected
    number of mistakes is at most $\eps\cdot T$. 
    This follows by considering 
    an hypothesis $h\in \H$ which realizes the branch corresponding to $(D_t,y_t)_{t=1}^T$.
    Indeed, for each fixed distribution $D_t$ on the branch $\mathcal{B}$, the probability that $h$ errs on $x_t\sim D_t$ is at most $\epsilon$.    
    Thus, by linearity of expectation, 
    the expected number of mistakes is at most $\eps\cdot T$, 
    and so there exists $h$ as stated.
\end{proof}

\begin{proof}[First proof of Theorem \ref{theorem-1}.]  This is immediate from Theorem \ref{thm:regret}, that is, 
  just as in the earlier proof of excellence, such a bound means that the set of elements in least one leaf cannot be split in a balanced way by any 
    $\epsilon$-good set, so must be $\epsilon$-excellent.\footnote{{In fact, since our quantification is over $\epsilon$-good distributions and 
    not just $\epsilon$-good sets, it is excellent in an a priori stronger sense than in the earlier proof. 
    We have not yet investigated how much stronger; as noted in \ref{d:motiv}, the move to distributions has other conceptual advantages.}}  
\end{proof}

\begin{disc}
To summarize and explain the use of Littlestone dimension hidden in this argument we emphasize that the above proof relies on deep ideas from online learning which are worth highlighting.  We also emphasize that this discussion surveys much prior work and not only what we do here. 

There are two background ideas: the first one has to do with no-regret algorithms (such as the multiplicative-weights algorithm, see e.g.~\cite{LittlestoneW94}).
Consider a set of $m<\infty$ experts (say weather forecasters), and every evening each of them tells us whether they think it will rain tomorrow or not.
Then, we use this list of predictions to make a prediction of our own.
Can we come up with a strategy that will guarantee that over $T$ days our prediction will not be much worse than that of the best expert in hindsight?
No-regret algorithms address this problem and provide {\it regret bounds} of roughly $\sqrt{T\cdot\log m}$, i.e. such algorithms make at most roughly $\sqrt{T\cdot\log m}$ mistakes more than the best of the $m$ experts in hindsight.
We stress that the $m$ experts can be arbitrary algorithms; in particular their prediction at day $t$ may be based 
on all information up to that point (i.e.\ prefix-dependence).

The second background idea, which is how the Littlestone dimension arises in the derivation of equation $(\ref{eq-ldim})$, 
is that any (possibly infinite) Littlestone class $\mch$
can be covered by a finite set of experts: that is, for every $T<\infty$ there is a set of $T$ choose $\leq \ldim(\mch)$
dynamic sets which simulate all hypotheses on $\mch$ with respect to \emph{sequences}\footnote{Below, we extend this to \emph{trees} of height $T$.} 
of length $T$.
Thus, by applying no-regret algorithms on the (finite!) set of experts we may ensure our regret is small relative to the best $h\in\mch$.

Together these explain existence of no-regret algorithms for any Littlestone class. 
Our bound on $d_\eps$, the height of an $\eps$-good complete shattered tree, 
exploits these connections in the new setting of $\eps$-good trees.
\end{disc}

\br

\section{Existence via closure properties} \label{s:closure}

In this section we give a second, quite different proof of Theorem \ref{theorem-1}. 
The resulting bound is significantly weaker than the one stated in Corollary~\ref{c:approxldim},
    but the reasoning may perhaps be more intuitive. In particular, it does not rely on the notion of regret from online learning.  
    The key ideas are the VC theorem and that classes remain Littlestone even after augmenting by fixed Boolean functions.  

\begin{rmk}
Continuing Discussion \ref{d:motiv}, note that these 
results can continue to make sense in the case where $\X, \H$ are uncountable but in that case inheriting the assumptions required 
to apply the VC theorem. 
\end{rmk}

{
\begin{rmk} The reader may choose to read this section either before or after \S \ref{s:ssp}. In the present order, 
Fact \ref{f:boolean} may be taken as a black box on a first reading, 
since the new proof of that fact we give below uses the results of $\S \ref{s:ssp}$. Still, 
the existence of a combinatorial companion proof of Theorem \ref{theorem-1} may be best appreciated 
in parallel to the proof just given, and the use of $\S \ref{s:ssp}$ in Fact \ref{f:boolean} may be a good motivation for 
those results for readers not familiar with definability of types. 
\end{rmk}
}

\begin{conv}
    In the rest of this section, let $\epsilon < \frac{1}{2}$ be arbitrary but fixed. 
\end{conv}

\begin{definition} Suppose we are given $k \in \mathbb{N}$ and some Boolean function $B:\{0,1\}^k\to \{0,1\}$. 
    Given $(X, \mch)$, let $(X, \mch^{(B)})$ denote the class
    \[\mch^{(B)} = \Bigl\{B(h_1,\ldots, h_k) : h_i\in \mch\Bigr\},\]
    where $B(h_1,\ldots, h_k)$ denotes the function which takes $x\in X$ to $B(h_1(x),\ldots h_k(x))\in\{0,1\}$.
\end{definition}

    Informally, we enrich $\mch$ by adding some additional hypotheses which come from applying our fixed $B$ 
    to $k$-tuples of elements of $\mch$.  (For example, we could start with $\mch$ which is a set of subsets of $X$, and 
    move to consider the hypothesis class whose elements are 
    intersections of pairs of elements of $\mch$; since we can recover any element of $\mch$ as its intersection with itself, 
    this is a kind of enrichment of $\mch.$)
    Observe that if the Boolean function sends the constant-$1$ sequence to $1$ and the constant-$0$ sequence to $0$, then 
    $\mch^{(B)} \supseteq \mch$. 
    
    We stress that although $B$ can be arbitrary, it is fixed for any instance of this construction.

{We will need the following fact. It was {proven} by \cite{ghazi21} (improving upon a previous bound by~\cite{alon20}), but 
we sketch below a new proof using our techniques.}

\begin{fact}[\cite{ghazi21}, Proposition 3] \label{f:boolean}
    If $\mch$ is a Littlestone class $($i.e.\ $\ldim(X, \mch)<\infty$$)$
    then also $\mch^{(B)}$ is a Littlestone class, and 
    \[\ldim\bigl(X, \mch^{(B)} \bigr) = O\Bigl(\ldim(X, \mch)\cdot k \cdot\log k\Bigr),\]
    where the big oh notation conceals a universal numerical constant.
\end{fact}

    Model theorists may check their intuition against the assertion that if $\vp$ is stable and $\psi$ is a fixed finite Boolean 
    combination of instances of $\vp$, then $\psi$ is also stable. 
    
    \begin{proof}[Proof Sketch] Let us sketch a proof of Fact \ref{f:boolean} using the language of \S \ref{s:ssp} below (a derivation using dynamic sets has not appeared in the literature, and is intuitive and analogous to the corresponding fact for VC classes). 
    If $\ldim(X, \mch) = d$ then for any integer $T$ we have a set $\mathcal{E}_T$ of 
    $\binom{T}{\leq d}$ dynamic sets which simulate $\mch$ on any $X$-labeled binary tree of height~$T$. To see that $\mch^{(B)}$ is also a 
    Littlestone class it would suffice to show that the same is true for some $d^\prime$ replacing $d$. 
    For each $T$, and for each $k$-tuple of dynamic sets 
    $E_1, \dots, E_k$ from $\mathcal{E}_T$, let $B(E_1, \dots, E_k)$ denote the dynamic set which operates by applying $B$ to the outputs of 
    $E_1, \dots, E_k$. Let $\mathcal{E}_T(B) = \{ B(E_1, \dots, E_k) : E_1, \dots, E_k \in \mathcal{E}_T \}$. Observe that this collection of dynamic sets 
    simulates $\mch^{(B)}$ on any $X$-labeled binary tree of height $T$ and its size will remain polynomial in $T$ (at most roughly $T^{dk}$).
    On the other hand, had $\mch^{(B)}$ not been Littlestone then one would need $2^T >> T^{dk}$ dynamic sets to cover it. 
    \end{proof}

    With Fact \ref{f:boolean} in hand, there is one more step: note we may also apply $B$ dually to $X$ rather than $\mch$.  
    To make sense of this, consider $(X, \mch)$ as a bipartite graph 
    with an edge between $x \in X$ and $h \in \mch$ if $h(x) = 1$.  In this picture, $\mch^{(B)}$ added some new points to the side of $\mch$ and 
    defined a rule for putting an edge between any such new point and any given element of $X$.  To apply $B$ dually, we carry out the 
    parallel operation for $X$ instead. 
    That is, let $(X^{(B)}, \mch)$ be the class where $X$ is enriched by new elements as follows: for any $x_1, \dots, x_k \in X$ define an element 
    $B(x_1,\dots, x_k)$ and for any $h \in \mch$, define $h(B(x_1, \dots, x_k)) = 1$ if and only if $B(h(x_1), \dots, h(x_k)) = 1$. 

    Recalling \ref{c:equiv}, the dual of a Littlestone class is a Littlestone class, so $(X^{(B)}, \mch)$ is Littlestone, though the $\ldim$ may be quite a bit larger.\footnote{It is known that $\ldim(X,\mch)\leq 2^{2^{\ldim(\mch,X)}}$ by applying Hodges' bound twice, see \S \ref{s:open-problems} \# \ref{hodges-bounds}.}

   \begin{concl} \label{c:bool}
   For any $k \in \mathbb{N}$ and any function $B: \{ 0, 1 \}^k \rightarrow \{ 0, 1 \}$, 
    if $(X, \mch)$ is a Littlestone class, then $(X^{(B)}, \mch)$ is a Littlestone class too.  
   \end{concl}

\begin{proof}[Second proof of Theorem \ref{theorem-1}]        
Suppose we are given an $\epsilon$-good tree~$\mct$ which is shattered by $\mch$. 
Choose $k$ large enough: $k=O(\mathsf{VCdim}(X, \mch)/(\frac{1}{2}-\eps)^2)$ will suffice. 
Let $B: \{ 0, 1 \}^k \rightarrow \{ 0, 1 \}$ be the 
majority vote operation given by $(x_1, \dots, x_k) \mapsto 0$ if $\{ 1\leq i \leq k : a_i = 0 \} \geq \frac{1}{2} k$ 
and $(x_1, \dots, x_{k}) \mapsto 1$ otherwise.
Suppose we independently sample $k$ elements $x_1, \dots, x_k$ 
from one of the $\epsilon$-good distributions labeling our given tree. Then by our choice of $k$, the VC theorem tells us 
that, with positive probability, the trace of each $h \in \mch$ on this sample is close enough to its true proportion. 
Here ``close enough'' means that the error is less than $\frac{1}{2}-\epsilon$. In particular, 
with positive probability, for \underline{every} $h \in \mch$ the majority vote $B(h(x_1), \dots, h(x_k))$ on this sample agrees with the opinion of
the $\epsilon$-good distribution on $h$.   We can therefore sample $k$ elements from each of the distributions labeling the nodes of the tree 
and with positive probability, all samples will be correct in this way. 

The crucial point is now that any full binary $\eps$-good tree $\T$ which is shattered by $\H$
    can be transformed to a (standard) full binary tree $\T'$ of the same height which is shattered by the class $(X^{(B)}, \mch)$.  
That is, there exists a choice of $k$ elements for each node in $\T$ such that the corresponding tree $\T'$ whose nodes are labelled by 
the $k$-wise majority votes of these elements (i.e. by the appropriate $B(x_1, \dots, x_k$)) is shattered by $(X^{(B)}, \mch)$.  
(To emphasize, in our Boolean-augmented class, these majority votes are represented by actual elements, and that is how the 
tree becomes a standard tree.) 
This shows that the length of $\T'$ (and also of $\T$) is bounded by $\ldim(X^{(B)}, \mch)$ which is finite by Conclusion \ref{c:bool}.  
(Note that this argument implicitly gives an inequality between the approximate and virtual Littlestone dimension, which are defined elsewhere.)
\end{proof} 

This argument is closer in spirit to similar arguments in VC theory concerning the variability of the VC dimension under natural operations. The obtained bounds however are much weaker than those of the previous section (at least double-exponentially weaker than the bound in Corollary~\ref{c:approxldim}).

\br

\section{Dynamic Sauer-Shelah-Perles lemmas for Littlestone classes} \label{s:ssp}

This section states and proves a mild variant of the celebrated Sauer-Shelah-Perles (SSP) lemma~\cite{sauer} 
replacing ``sequences of length $T$'' by ``trees of height $T$'' (informally, the adversary can change the elements 
we are given in response to our past choices).  
This should also allow the model-theoretic reader to understand the key use of Littlestone dimension in \S \ref{s:e-r}, where the case of 
sequences was already sufficient.

    Let $\mch$ be a class with Littlestone dimension $d<\infty$.
    Two results which could be considered variants of the SSP lemma are known for Littlestone classes:
    the first one, observed by Bhaskar~\cite{bhaskar} provides an upper bound of $T$ choose $\leq d$ 
    on the number
    of leaves in a binary-tree of height $T$ with $X$-labeled nodes that are reachable by $\H$.
    The second, \emph{dynamic} version is due to Ben-David, Pal, and Shalev-Shwartz~\cite{ben-david09agnostic}.
    This lemma is a key ingredient in the characterization of (agnostic) online-learnability by Littlestone dimension;
    it asserts the existence of $T$ choose $\leq d$  
    online algorithms (or experts or dynamic-sets)
    such that for every sequence $x_1,\ldots, x_T$ and for every $h\in \mch$
    there exists an algorithm among the ${T \choose \leq d}$ algorithms
    which produces the labels $h(x_1),\ldots , h(x_T)$ when given the sequence $x_1,\ldots, x_T$
    as input.  Again, the version we shall prove is a mild extension of the 
    Ben-David, Pal, Shalev-Shwartz lemma to the case of trees rather than sequences. 
    We shall give the statement, then present the key terms, then give the proof.

\begin{theorem} \label{dssp} Let $(X, \mch)$ be a Littlestone class of dimension $d$. 
For every $T \in \mathbb{N}$ there exists a collection $\mathbf{A}$ of $\binom{T}{\leq d}$ algorithms $($dynamic sets$)$ such that for every binary tree 
$\mct$ of height $T$ with $X$-labeled internal nodes, every branch in $\mct$ which is realized by some $h \in \mch$ is also realized by some 
algorithm from $\mathbf{A}$. 
\end{theorem}

\begin{rmk}
{For simplicity, we define dynamic sets to be deterministic, but it is also reasonable for them to be random. 
In general, a randomized algorithm is simply a distribution over deterministic algorithms. When a randomized algorithm is a distribution over prefix-dependent deterministic algorithms, see below, then we may say it is prefix-dependent.}
\end{rmk}

\begin{definition}[In the language of online learning]  Fix  $~T \in \mathbb{N}$ and a set $X$. 
A dynamic set (or adaptive expert) $\mathcal{A}$ is a function which assigns to each internal node in each $X$-labeled binary tree of height $\leq T$ a value in $\{ 0, 1 \}$ in a prefix-dependent way. 
\end{definition}

To explain, notice that $\mathcal{A}$ naturally defines a walk in any such tree: it starts at the root which is labeled by 
some $a_\emptyset =: a_0$, it outputs $\mathcal{A}( \langle a_0 \rangle) =: t_0$, then travels left (if its output was $0$) or right (if its output was $1$) to a node 
labeled by $a_{\langle t_0 \rangle} =: a_1$, where it outputs  $\mathcal{A}(\langle a_0, a_1 \rangle ) =: t_1$ and so on. 
Prefix-dependence means that for any $\ell \leq T$, if in two different trees the sequences of values $a_0, \dots, a_\ell$ produced 
in this way are the same, then also the output $t_\ell$ of $\mathcal{A}$ in both cases is the same. 

It should now be clear what it means for an algorithm $\mathcal{A}$ to realize a branch in a tree (the directions it gives  
instruct us to walk along this root-to-leaf path). Note that we can think of each $h \in \mch$ as a very simple dynamic set in 
its guise as a characteristic function.  

\begin{rmk}
In online learning one distinguishes between adaptive and oblivious experts (or between experts with and without memory): an oblivious expert is simply an $X\to\{0,1\}$ function, 
whereas an adaptive expert has memory and can change its prediction based on previous observations. 
The above definition captures adaptive experts. The above definition slightly deviates from the standard definition of adaptive experts. In the standard definition, one usually only considers sequences (or oblivious trees), rather than general trees. Notice that the distinction between oblivious and general trees can be expressed analogously with respect to the adversary: the adversary, who presents the examples to the online learner, can be oblivious -- in which case it decided on the sequence of examples in advance, or it can be adaptive -- in which case it decides which example to present at time $t$ based on the predictions the online learning algorithm made up to time $t$. In this language, our version of the dynamic SSP applies also to adaptive adversaries, whereas the previous version was restricted to oblivious adversaries. 
\end{rmk}

\begin{definition}[In more set-theoretic language]
Let $T \in \mathbb{N}$, and $\kappa = |X|^T$. 
Consider the set $\mathcal{E} = \langle e_i : i < \kappa \rangle$ of all $T$-element sequences of elements of $X$. 
A \emph{dynamic set} assigns to each enumeration $e_i$ a function $f_i : T \rightarrow \{ 0, 1 \}$, and the assignment must be 
\emph{coherent} in the sense that if $e_i \rstr \beta = e_j \rstr \beta$ then $f_i \rstr \beta = f_j \rstr \beta$. 
\end{definition}

\begin{expl}
Let $X = \mathbb{N}$. Let $\mathcal{A}$ be the algorithm which receives $a_t$ at time $t$ and 
outputs $1$ if $a_t$ is the largest prime it has seen so far and $0$ otherwise. 
\end{expl}

\begin{definition}
Given a possibly partial characteristic function $g$ with $\dom(g) \subseteq X$ and $\operatorname{range}(g) \subseteq \{ 0, 1 \}$, define the 
``version space'' $H_g = \{ h \in \mch : g \subseteq h \}$.
\end{definition}

\begin{rmk} \label{l:half}
Observe that if $\mch$ is a Littlestone class, 
$H \subseteq \mch$ and $f$ is a possibly partial, possibly empty characteristic function with $\dom(f) \subseteq X$ and $a \in X$, then 
\[ (\star) ~~~\min \{ \ldim(H_{f \cup \{ (a, 0) \}}), \ldim(H_{f \cup \{ (a, 1) \}}) \} < \ldim (H_{f}) \]
i.e., on one side of any partition by a half-space the dimension must drop, 
since $\ldim(H_{f}) \leq \ldim(\mch) = d$ is defined and finite. $($This property is what enables the notion of Littlestone majority vote.$)$
\end{rmk}

\begin{proof}[Proof of Theorem \ref{dssp}]
To define our algorithms some notation will be useful.  By ``tree'' in this proof we always mean an $X$-labeled binary tree of height $T$. 
Given an algorithm $\ma$ and a tree $\mct$, let $\sigma = \sigma(\ma, \mct) = \langle a_i : i < T \rangle$ 
and let $\tau = \tau(\ma, \mct) = \langle t_i : i < T \rangle$ denote the sequence of elements of $X$ associated to the nodes traversed, 
and the corresponding outputs of $\ma$, respectively. Again, given $\ma$ and $\mct$, let $\gamma = \gamma(\ma, \mct) = \langle g_i : i < T \rangle$ be the sequence of partial characteristic functions 
given by $g_i = \{ (a_j, t_j) : j < i \}$.

We define the $\binom{T}{\leq d}$ algorithms as follows. Each algorithm $\ma$ is parametrized by a set $A \subseteq\{ 0, \dots, T-1 \}$ of size $\leq d$ 
(and there is an algorithm for each such set). 
Given any tree $\mct$, the algorithm proceeds as follows. Upon reaching a node at level $i$ labeled by $a_i$, it computes the values 
$\ldim(H_{g_i \cup \{ (a_i, 0) \}})$ and $\ldim(H_{g_i \cup \{ (a_i, 1) \}})$.   
Informally, it asks how the Littlestone dimension of the set 
$H_{g_i}$ will change according to the decision on~$a_i$. It then makes its decision by cases. 
If $i \in A$, then the algorithm chooses the value of $t_i$ which will make $\ldim(H_{g_i \cup \{ (a_i, t_i) \}})$ smaller, and in case of ties chooses $0$. 
If $i \notin A$, then the algorithm chooses the value of $t_i$ which will make $\ldim(H_{g_i \cup \{ (a_i, t_i) \}})$ larger, and in case of ties chooses~$1$.  
This finishes the definition of our class $\mathbf{A}$.  Clearly the algorithms involved are all prefix dependent. 

Let us verify that for any tree $\mct$ and any $h \in \mch$ there is an algorithm in $\mathbf{A}$ realizing the same branch as $h$. 
Let $(b_0, s_0), \dots, (b_{t-1}, s_{t-1})$ denote the root-to-leaf path traversed by $h$.  For each $i < T$, let 
$f_i = \{ (b_j, s_j) : j < i \}$ denote the partial characteristic function in play as we arrive to $b_i$.  
(Notice that necessarily each $f_i \subseteq h$.)  
Let us consider how we may use $A$ to signal what to do.
Let $d^i = \ldim(H_{f_i})$, 
let $d^i_0 = \ldim (H_{f_i \cup \{ (a_i, 0) \}}$ and let 
$d^i_1 = \ldim (H_{f_i \cup \{ (a_i, 1) \}}$. 
There are several cases. If we know that at stage $i$ the $\ldim$ 
does not drop then by \ref{l:half} the choice is determined. 
If we know that the $\ldim$ drops and $d^i_0 \neq d^i_1$ then the 
choice is determined by knowing whether we chose the larger or smaller.  If we know that $\ldim$ drops and $d^i_0 = d^i_1$ then the 
choice is determined by knowing whether or not we went left. With this in mind, 
define $B \subseteq \{ 0, \dots, T-1 \}$ to be $B = \{ i < T : 
( ~\ldim(H_{f_i}) \geq \ldim (H_{f_i \cup \{ (a_i, 1-s_i) \}}) 
> \ldim (H_{f_i \cup \{ (a_i, s_i) \}})~) $ or $ (\ldim(H_{f_i \cup \{ (a_i, s_i) \}}) = \ldim((H_{f_i \cup \{ (a_i, 1-s_i) \}})  = \ldim(H_{f_i})-1$ and 
$s_i = 1 ) \}$. 
In English, $B$ is the set of all $i < T$ at which either there was only one way to make the dimension drop as much as possible, 
or both ways the dimension dropped by the same amount and we went left.  Since at every $i \in B$ the Littlestone dimension drops, necessarily $|B| \leq d$. 

Consider the algorithm $\ma \in \mathbf{A}$ parameterized by $B$. We argue by induction on $i<T$ that 
$g_i = f_i$, that is, $a_i = b_i$ and $t_i = s_i$.  To start, $a_0 = b_0$ is the label of the root.  
If $i \notin B$, then at this stage along the path traversed by $h$, either the 
Littlestone dimension did not drop as much as possible or $s_i =1$. 
In the first case, 
there is only one value of $t_i$ which will keep the dimension larger, 
and that is $t_i = s_i$.  If the dimension went down equally for both successors, $\ma$ will choose $t_i = 1 = s_i$.  
If $i \in B$, then here the Littlestone dimension must drop as much as possible, so either there is only one way to achieve this and 
so $t_i = s_i$, or both successors drop equally and 
$t_i = 0 = s_i$.  This completes the proof. 
\end{proof}

\begin{rmk}
A model theoretic reader will see definability of $\vp$-types for stable $\vp$. 
\end{rmk}

We now verify that this is a characterization.  

\begin{lemma}
Suppose $\ldim(X, \mch)$ is not finite. 
Then for every $d \in \mathbb{N}$, for all sufficiently large $T \in \mathbb{N}$ 
and every collection $\mathbf{A}$ of $\binom{T}{\leq d}$ dynamic sets, there is some binary tree $\mct$ of height $T$ with 
$X$-labeled internal nodes and some $h \in \mch$ which realizes a branch in $\mct$ not realized by any algorithm from $\mathbf{A}$. 
\end{lemma}

\begin{proof} 
Choose $T$ so that $2^T > T^d$. 
Since $\ldim$ is not finite, we may construct a full binary tree of height $T$ whose nodes are labeled by $X$ and such that every branch is 
realized by some $h \in \mch$.  Every algorithm $\ma \in \mathbf{A}$ realizes one and only one branch in $\mct$, so there are 
not enough of them to cover all branches.
\end{proof}

To conclude, observe $\mathbf{A}$ simulates $\mch$ in an even stronger way: 
its algorithms can continue to simulate the realization of branches 
by $\mch$ even when we weaken the notion of realization to allow a certain number of mistakes.  (This also gives a simple derivation of the 
``oblivious'' SSP lemma from our ``tree'' version here.)

\begin{corollary} \label{c:path} 
Suppose $\ldim(\mch) = d \in \mathbb{N}$, 
let $T \in \mathbb{N}$, and let $\mathbf{A}$ be the family of $\binom{T}{\leq d}$ algorithms 
constructed for $\mch$ in Theorem $\ref{dssp}$.  Let $\mct$ be any binary tree of height $T$ with $X$-labeled internal nodes. Given any  
branch $(a_0, t_0), \dots, (a_{T-1}, t_{T-1})$ and any $h \in \mch$, let $S = \{ i < T : h(a_i) \neq t_i \}$ be the set of ``mistakes'' made by 
$h$ for this branch. Then there is $\ma \in \mathbf{A}$ which makes the same set of mistakes for this branch.  
\end{corollary}

\begin{proof}
Consider a new tree $\mct_*$ where all the nodes at level $i$ have the same label $a_i$ (the tree is ``oblivious'').  
So branches through $\mct_*$ amount to choosing subsets of $\{ a_i : i < T \}$. This particular tree also 
falls under the jurisdiction of Theorem \ref{dssp}, and so gives our corollary. 
\end{proof}

\br

\section{Majorities in Littlestone classes}

So far we have been guided by the thesis that Littlestone classes are characterized by frequent, large sets 
with well-defined notions of majority. However, there are at least two candidate notions of majority which are quite distinct: 
the majority arising from the counting measure, which we have been exploring via $\epsilon$-excellent and $\epsilon$-good, and the notion of 
majority arising from Littlestone rank. 

In this section we prove that these two notions of majority ``densely often agree'' in Littlestone classes 
and indeed this is true of any simple axiomatic notion of majority, as defined below. 

\begin{definition} 
Say that $\textcolor{red}{H} \subseteq \mch$ is \emph{Littlestone-opinionated} if for any $a \in X$, one and only one of 
\[ \ldim(\{  h \in H : h(a) = 0 \}),   \ldim(\{ h \in H : h(a) = 1 \})   \]
is strictly smaller than $\ldim(\textcolor{red}{H})$. 
\end{definition}

As a warm-up, we prove several claims.  
As above, a \emph{partition of $H$ by a half-space}
means that for some element 
$a \in X$ we separate $H$ into $\{ h \in H : h(a) = 0 \}$ and $\{ h \in H : h(a) = 1 \}$.)
{So in this language, $H$ is Littlestone-opinionated if in any partition by a half-space, exactly one of the two 
pieces retains the $\ldim$.}

\begin{claim} \label{c:42}
Suppose $\ldim(\mch) = d$ and $0 < \epsilon < \frac{1}{2}$. Then for any finite 
$H \subseteq \mch$ there is $A \subseteq H$ of size $\geq \epsilon^d|H|$ such that $A$ is both $\epsilon$-good and Littlestone-opinionated and 
these two notions of majority agree, i.e. for any $a \in X$ 
\[ \ldim(\{ h \in A : h(a) = t \}) = \ldim( \textcolor{red}{A} )  \mbox{ iff } | \{ h \in A : h(a) = t \} | \geq (1-\epsilon)|A|.
\]
\end{claim}

\begin{proof}
It suffices to observe that given any finite $H \subseteq \mch$ which (a) is not $\epsilon$-good, (b) is $\epsilon$-good but is not Littlestone-opinionated, or 
(c) is both $\epsilon$-good and Littlestone-opinionated but the two notions of majority do not always agree, we can find 
$G \subseteq H$ (arising from by a partition of $H$ by a half-space) 
with $|G| \geq \epsilon |H|$ and $\ldim(G) < \ldim(H)$, because this initiates a recursion which cannot continue 
more than $\ldim(H) \leq \ldim(\mch) = d$ steps.  

Why? In case (a), there is a partition into two pieces of size $\geq \epsilon|H|$; choose the one of smaller $\ldim$. 
In case (b), there is a partition into two pieces each of $\ldim$ strictly smaller than $\ldim(H)$; choose the one of larger counting measure. 
In case (c), there is a partition where the majorities disagree, and we can choose the piece of larger counting measure and thus smaller 
$\ldim$. This completes the proof. 
\end{proof}

\begin{definition} \label{d:gp}
Call $P$ a \emph{good property} $($or: $\epsilon$-good property$)$ for $\mch$ 
if it is a property of finite subsets of $\mch$ which implies $\epsilon$-good and which satisfies: for some constant 
$c = c(P) > 0$, for any finite $H \subseteq \mch$ there is $B \subseteq H$ of size $\geq \epsilon^c|H|$ with property $P$. 
\end{definition}

\begin{corollary}
By $\ref{c:one-half}$ above, if $\mch$ is Littlestone and $\epsilon < \frac{1}{2}$ then 
 ``$\epsilon$-excellent'' is an $\epsilon$-good property for $\mch$. 
\end{corollary}

\begin{lemma} \label{lemma:ep}
Let $\mch$ be a Littlestone class of dimension $d$ and $0 < \epsilon < \frac{1}{2}$.  Let $P$ be a good property for $\mch$ and $c = c(P)$. 
Then for any finite $H \subseteq \mch$ there is $A \subseteq H$ of size $\geq \epsilon^{(c+1)d}|H|$ such that $A$ has property $P$ 
$($so is also $\epsilon$-good$)$ and is Littlestone-opinionated, and for any $a \in X$, 
\[ \ldim(\{ h \in A : h(a) = t \}) = \ldim( H )  \mbox{ iff } | \{ h \in A : h(a) = t \} | \geq (1-\epsilon)|A| \]
i.e. the $\epsilon$-good majority and the Littlestone majority agree. 
\end{lemma}

\begin{proof}
Modify the recursion in the previous proof as follows. At a given step, if $H$ does not have property $P$,  
replace it by a subset $C$ of size $\geq \epsilon^c|H|$ which does. 
Since $P$ implies $\epsilon$-good, if we are not finished, then we are necessarily in case (b) or (c) 
and at the cost of an additional factor of $\epsilon$ we can find $B \subseteq C$ where the $\ldim$ drops.  
In each such round, we replace $H$ by $B \subseteq H$ with $|B| \geq \epsilon^{c+1} |H|$ and $\ldim(B) < \ldim(H)$. 
\end{proof}

\begin{disc}
\emph{Note that this majority agreement deals with half-spaces, which is arguably the 
interesting case for ``Littlestone-opinionated'' as it relates to the SOA.  
In \ref{lemma:ep}, it is a priori not asserted that every subset of $A$ which is large in the sense of counting measure (but does not 
arise from a half-space) has large $\ldim$.} 
\end{disc}

\begin{definition}[Axiomatic largeness]  \label{d:arl}
Define $\mcm$ to be an axiomatic notion of relative largeness for the class $\mch$ if it satisfies the following properties.\footnote{{This captures  \emph{relative} majority or largeness since ``being large in'' is a two-place relation.}}
\begin{enumerate}
\item $\mcm$ is a subset of $\mcp = \{ (B, A) :  B \subseteq A \subseteq \mch \}$. 

\item Define $\mcp_{half} := \{ (B, A) \in \mcp : B$ arises as the intersection of $A$ with a half-space $\}$.

\item If $(B, A) \in \mcm$, say ``$B$ is a large subset of $A$.''  We may write $B \subseteq_\mcm A$.

\item The rules are: 
\begin{enumerate}

\item {\emph{(monotonicity in the set)} if $C \subseteq B \subseteq A$ and $C \subseteq_\mcm A$ then $B \subseteq_\mcm A$.}

\item {\emph{(monotonicity in the superset)} if $C \subseteq B \subseteq A$ and 
$C \subseteq_\mcm A$ then $C \subseteq_\mcm B$.}

\item \emph{(identity)} $(A, A) \in \mcm$. 

\item \emph{(non-contradiction)} If $(B, A) \in \mcp_{half}$ and $C = A \setminus B$ [so also $(C, A) \in \mcp_{half}$] 
then at most one of $(B, A)$ and $(C, A)$ belongs to $\mcm$. 

\item \emph{(chain condition)} There is $n = n(\mcm) < \omega$ such that if $\langle A_i : i < m \rangle$ is a set of subsets of $\mch$ and $(A_{i+1}, A_{i}) \notin \mcm$ for all $i < m-2$ then $m \leq n$.  In other words, the length of a descending chain 
\[ A_{m-1} \subseteq  \cdots \cdots \subseteq A_0 \] 
of non-large subsets is upper bounded by $n$. 
\end{enumerate}
\end{enumerate}
\end{definition}

\begin{expl} \label{e:11a}
Suppose $\mch$ is a Littlestone class. Then 
\[ \mcm = \{ (B, A) \in S : \ldim(A) = \ldim(B) \} \]
satisfies Definition $\ref{d:arl}$. 
\end{expl}

\begin{proof}
Conditions (3)(a),(b),(c) are immediate; (d) follows by the definition of $\ldim$. 
Condition (e) is clear because if $(A_i, A_{i+1}) \notin \mcm$ then 
$\ldim(A_{i+1}) < \ldim(A_i)$ so $n(\mcm) \leq d$. 
\end{proof}

\begin{expl}
{For model theoretic readers, note that for suitable hypotheses classes the Shelah $R(x=x, \Delta, \lambda)$ ranks, 
when defined and restricted to multiplicity one, can illustrate $\ref{d:arl}$ for other values of 
$\Delta$, $\lambda$.} 
\end{expl}

In \ref{l:majority-b} we don't need to assume a priori that $\mch$ is Littlestone, though the proof will show that it is. 

\begin{lemma} \label{l:majority-b}
Suppose $\mch$ admits a notion of relative largeness $\mcm$. Let $0 < \epsilon < \frac{1}{2}$ and let 
$P$ be an $\epsilon$-good property for $\mch$. Let $c = c(P)$ and $n = n(\mcm)$. Then for any nonempty finite 
$H \subseteq \mch$ there is $A \subseteq H$ of size $\geq \epsilon^{(c+1)n}|H|$ such that:
\begin{enumerate}
\item $A$ has property $P$, and thus is $\epsilon$-good, so for any partition of $A$ by a half-space into $B \cup C$, at least one 
$($so exactly one$)$ of $B, C$ has size $<\epsilon|A|$.

\item For any partition of $A$ by a half-space into $B \cup C$, at least one $($so exactly one$)$ of $(B, A)$, $(C, A)$ belongs to $\mcm$. 

\item The two notions agree, i.e. $(B, A) \in \mcm$ if and only if $|B| \geq (1-\epsilon)|A|$. 
\end{enumerate}
\end{lemma}

\begin{proof}
Let $n = n(\mcm)$ and set $A_0 = H$. By induction on $t \geq {0}$ we shall prove that if 
$A_t$ does not satisfy contitions 1, 2, and 3 then either it contains a subset of size $\geq \epsilon^c|A_t|$ which does, or 
there is $A_{t+1} \subseteq A_t$ such that 
$|A_{t+1}| \geq \epsilon^{c+1}|A_t|$ and $(A_{t+1}, A_t) \notin \mcm$.  
Our chain condition \ref{d:arl}(4)(e) will then ensure $t \leq n$.  

For each $t \geq 0$ proceed as follows.  
If $A_{t}$ has property $P$, define $A^\prime_t = A_t$. 
If not, replace $A_t$ by a subset of size $\geq \epsilon^c|A_t|$ which does, and set this to be $A^\prime_{t}$. 
A priori, we have no information on whether $(A^\prime_t, A_t) \in \mcm$. 
Since $A^\prime_t$ has property $P$, if $A^\prime_t$ does not already satisfy 1, 2, and 3, then condition 2 or 3 must fail; in either case, there must be some half-space which 
partitions $A^\prime_t$ into two non-trivial sets at least one of which, call it $B$, has size at least $\epsilon|A^\prime_t|$ and satisfies 
$(B, A^\prime_t) \notin \mcm$. Set $A_{t+1} = B$. Then $|B| \geq \epsilon^{c+1}|A_t|$ and by condition \ref{d:arl}(4)(b), $(A_{t+1}, A_t) \notin \mcm$.  
This completes the inductive step and the proof. 
\end{proof}

\begin{theorem}  \label{theorem-3} 
The following are equivalent for $(X, \mch)$.

\begin{enumerate}
\item $\mch$ admits a notion of relative largeness $\mcm$.

\item $\mch$ is a Littlestone class. 

\item For every $\mcm$ and $0 < \epsilon < \frac{1}{2}$ there is $n = n(\epsilon, \mcm)$ such that every finite nonempty $H \subseteq \mch$ 
has a subset $A$ which satisfies:

\begin{enumerate}

\item $|A| \geq \epsilon^n|H|$, and

\item $A$ is $\epsilon$-good, and 

\item for every partition of $A$ by a half-space into $B \cup C$, $(B, A) \in \mcm$ if and only if $|B| \geq \epsilon|A|$ if and only if 
$|B| \geq (1-\epsilon)|A|$. 

\end{enumerate}
i.e. the counting majority and the $\mcm$-majority are well defined and agree.
\item In item (3) we may replace $(b)$ by ``$A$ has property $P$'' 
when $P$ is an $\epsilon$-good property for $\mch$, at the cost of changing the exponent $n$ in $(a)$ to $(c+1)n$ for $c = c(P)$.$)$
\end{enumerate}
\end{theorem}

\begin{proof}
(2) implies (1) is Example \ref{e:11a}. (1) implies (3) [or (4)] is Lemma \ref{l:majority-b}. Clearly (4) implies (3). For (3) implies (2), note that (3) 
tells us a fortiori that we can always find large $\epsilon$-good subsets, so $\mch$ must be a Littlestone class by \ref{c:equiv}. 
\end{proof}

\begin{rmk}
Although we have formulated these largeness properties for subsets of $\mch$, the symmetric results whould hold for subsets of $X$. 
\end{rmk}

\begin{disc} 
\emph{A key point in the proof of the stable regularity lemma (in our language) is that 
because finite Littlestone dimension implies finite VC dimension, if we randomly partition $\epsilon$-excellent sets {then} the pieces 
are likely to remain excellent (for a related $\epsilon$). This is what allowed for an equitable partition into excellent sets. 
There is a priori no reason the analogous fact should be true for Littlestone-opinionated sets. However, this section finds (densely often) sets where  
the counting majority and the Littlestone majority agree. Randomly partitioning these, we retain goodness, so necessarily also 
retain the ability to correctly predict Littlestone majority in accordance with the original set, despite perhaps not being  
Littlestone-opinionated.} 
\end{disc}

\begin{disc}
\emph{It is interesting to inspect relative largeness from the perspective of online learning.
    Indeed, note that any such notion gives rise to an online learning strategy with a bounded mistake bound:
    the online learner maintains a version space $\mch_i\subseteq \mch$, starting with $\mch_0=\mch$.
    For each input example $x_i$ received, the learner predicts $\hat y_i$ such that 
    \[ (\{h\in \mch_i : h(x_i) = \hat y_i\}, \mch_i)\in \mcm \]
    and note that there can be at most such $\hat y_i$, if no such $\hat y_i$ exists then the learner predicts $\hat y_i = 0$.
    Then, upon receiving the true label $y_i$, 
    if $y_i=\hat y_i$ then the learner sets $\mch_{i+1}=\mch_i$ and else,
    when $y_i\neq \hat y_i$, the learner sets $\mch_{i+1} = \{h\in \mch_i : h(x_i)=y_i\}$.
    Observe that given any sequence $(x_1,y_1),\ldots, (x_T,y_T)$, this learner makes at most $n(\mcm)$ mistakes:
    indeed, if the learner makes a mistake on $x_i$ then $(\mch_{i+1},\mch_{i})\notin\mcm$,
    and $\mch_{i+1}$ is obtained by intersecting $\mch$ with a halfspace.}

\emph{
This point of view offers an alternative explanation for the fact that only Littlestone classes admit notions of relative largeness.
    Moreover, it implies that for every notion of relative largeness $\mcm$, 
    we have that $n(\mcm)$ is at least the Littlestone dimension.
    This follows because the Littestone dimension is equal to the optimal mistake-bound.
    Thus, the notion of relative largeness which arises from the Littlestone dimension is
    optimal in the sense that it minimizes $n(\mcm)$.}
\end{disc}

\br

\section{Some open problems} \label{s:open-problems}

To conclude the paper we mention several natural open problems and directions 
for further work; {some appear challenging, some more accessible.} 

\begin{enumerate}
\item For VC classes, recall that we have the usual Sauer-Shelah-Perles lemma, and 
Haussler's covering lemma which says that every VC class of VC-dimension $d$ 
can be $\epsilon$-covered by roughly $\frac{1}{\epsilon^d}$ hypotheses~\cite{Haussler95}.\footnote{Informally, there is a list of approximately $\frac{1}{\epsilon^d}$ hypotheses 
such that every other hypothesis in our class is $\epsilon$-close to one 
of the hypotheses in our list.} (The SSP lemma can be thought of as the special case of Haussler's covering lemma where the domain has size $n$ and when $d = \frac{1}{n}$.)
This is clearly useful for learning. 
It is natural to ask whether there is a dynamic version of this 
covering lemma for Littlestone classes. That is, is there a function 
$f=f(\epsilon, d)$ such that for any Littlestone class $\mch$ of $\ldim$ $d$ 
we can always find $\leq f(\epsilon, d)$ dynamic sets which approximately
cover the whole class $\mch$, meaning that for every sequence $x_1, \dots, x_n$ 
from $X$ and every $h \in \mch$ there is a dynamic set in our list 
which is $\epsilon$-close to it. This is also related to \cite{alon21}.

\item \label{hodges-bounds}
In the classical case, there is a fundamental relationship between the 
Littlestone dimension and the half-graph/Threshold dimension, as 
explained by Shelah's unstable formula theorem: both are finite together, 
and bounds are known \cite{shelah}, \cite{hodges}. However, the question of determining 
tight quantitative bounds in the finite remains open.  {The known bounds given by Hodges \cite{hodges1} 
say roughly that from a half-graph of length $k$ there is a tree of height about $\log k$, and that 
from a tree of height $n$ there is a half-graph of length about $\log n$.} Determining whether 
these bounds are tight may be challenging, and seems worth emphasizing here. 
To reiterate what we said in section one, a useful 
aspect of this relationship is connecting a symmetric or self-dual 
quantity with Littlestone dimension (see 1.3 above). 

\item As mentioned in the text, applying the Hodges bounds twice tells us that if $\ldim(\mch) = d$, 
the dual class has Littlestone dimension bounded by about $2^{2^d}$. Can this be improved? 

\item It seems worth while 
to explore further the significance of dynamic Sauer-Shelah-Perles lemmas for model theory. 
As a soft question, are there useful model theoretic explanations for the existing regret bounds for multiplicative-weights algorithms 
mentioned in \S \ref{s:e-r}?

\item We have not sorted out the extent of the nonmonotonicity of excellence as $\epsilon$ varies; 
this could give another approach to existence of excellent sets. Nor have we  
tried to optimize the constants in either of the two existence proofs in the present paper.  

\item Is there a helpful ``outside'' characterization of the good properties in the sense of \ref{d:gp}?

\end{enumerate}

\br

\bibliographystyle{amsplain}



\end{document}